# UPDATE ON THE CODE INTERCOMPARISON AND BENCHMARK FOR MUON FLUENCE AND ABSORBED DOSE INDUCED BY AN 18 GEV ELECTRON BEAM AFTER MASSIVE IRON SHIELDING[†]

A. Fassò[1], A. Ferrari[2], A. Ferrari[3], N.V. Mokhov[4], S.E. Mueller[3,#], W.R. Nelson[1], S. Roesler[2], T. Sanami[5], S.I. Striganov[4], R. Versaci[6]

[1]SLAC National Accelerator Laboratory (retired), USA
[2]CERN, Switzerland
[3]Helmholtz-Zentrum Dresden-Rossendorf, Germany
[4]Fermi National Accelerator Laboratory, USA
[5]KEK, Japan
[6]ELI Beamlines, Czech Republic

## ABSTRACT

*In 1974, Nelson, Kase and Svensson published an experimental investigation on muon shielding around SLAC high-energy electron accelerators [1]. They measured muon fluence and absorbed dose induced by 14 and 18 GeV electron beams hitting a copper/water beamdump and attenuated in a thick steel shielding. In their paper, they compared the results with the theoretical models available at that time.*

*In order to compare their experimental results with present model calculations, we use the modern transport Monte Carlo codes MARS15, FLUKA2011 and GEANT4 to model the experimental setup and run simulations. The results are then compared between the codes, and with the SLAC data.*

---

# UPDATE ON THE CODE INTERCOMPARISON AND BENCHMARK FOR MUON FLUENCE AND ABSORBED DOSE INDUCED BY AN 18 GEV ELECTRON BEAM AFTER MASSIVE IRON SHIELDING

A. Fassò[1], A. Ferrari[2], A. Ferrari[3], N.V. Mokhov[4], S.E. Mueller[3,#], W.R. Nelson[1], S. Roesler[2], T. Sanami[5], S.I. Striganov[4], R. Versaci[6]

[1]SLAC National Accelerator Laboratory (retired), USA
[2]CERN, Switzerland
[3]Helmholtz-Zentrum Dresden-Rossendorf, Germany
[4]Fermi National Accelerator Laboratory, USA
[5]KEK, Japan
[6]ELI Beamlines, Czech Republic

**Introduction**

Since muons emit less bremsstrahlung radiation when passing through matter than electrons due to their larger mass (and therefore lose less energy along their way), adequate shielding design is required for future beamline facilities like ELI-Beamlines, LCLS at SLAC and the planned ILC in Japan. As an example, it has been demonstrated that photoproduced muons at ELI-Beamlines in Prague [2], with a 10 GeV electron beam for acceleration experiments may affect neighboring labs [3]. A good understanding of photoproduction of muon pairs by high-energy, high-intensity electron or gamma beams is therefore required. In order to estimate how well theoretical models for photoproduction of muons in current radiation transport codes describe the reality, we compare Monte Carlo calculations using the codes FLUKA2011, MARS15 and GEANT4 with data from an experiment done in 1974 at the Stanford Linear Accelerator Center in California. In this way we can also compare the results of the different codes with each other. While preliminary results have been already presented at the last SATIF workshop [4], this document gives an update on the current status of the work.

**The experiment**

In 1974, Nelson, Kase and Svensson carried out an experimental investigation at SLAC to study the muon fluence and absorbed dose induced by an 18 GeV electron beam hitting a copper/water beamdump [1]. In the vicinity of a nucleus, the electrons produced bremsstrahlung photons in the beamdump, which subsequently lead to muon pair photoproduction. The muons were produced within 6 radiation lengths in the beamdump (corresponding to 22.23 cm), and were subsequently attenuated by thick blocks of shielding iron. The lateral distribution of the muon fluence and the absorbed dose were measured by positioning detectors perpendicular to the incident electron beam axis in four narrow gaps (gap A, gap B, gap C, gap D) between the iron shielding blocks. The muon fluence was detected using 400μm thick nuclear track emulsion plates, which were read out by microscopes after the exposure. Thermoluminescent dosimeters were used to register the absorbed dose. In addition, two scintillation counters determined the exposure and also crosschecked the muon fluence measurements. The geometry of the setup allowed to perform measurements at vertical angles from 0 to 150 milliradians. Within this range, it is ensured that the direct flight paths from the muon production point in the beam dump to their detection are completely contained in the iron shielding. To protect the detectors against background radiation, the gaps A, B and C were covered with lead blocks on the side and on top. Gap D, which is the furthest away from the muon production point, was left exposed.

**Monte Carlo transport codes**

A first comparison of results from the transport codes MARS [5-9] and FLUKA [10,11] with the experimental results was done in 2007 [12]. Together with the observations made in the shielding design for the ELI beamlines facility in Prague, the decision was taken to redo the comparison with the newest versions of the two codes, and also include the GEANT4 [13,14] toolkit as a third transport code into the comparison.

*FLUKA2011*
FLUKA is a fully integrated particle physics Monte Carlo simulation package containing implementations of sound and modern physical models. A powerful graphical interface (FLAIR [15]) facilitates the editing of FLUKA input, execution of the code and visualization of the output. Photomuon production in FLUKA is implemented as coherent nuclear scattering according to the Bethe-Heitler mechanism using the formalism of [16,17].

*MARS15*
The MARS code is a general-purpose, all-particle Monte Carlo simulation code. It contains established theoretical models for strong, weak and electromagnetic interactions of hadrons, heavy ions and leptons. Most processes can be treated exclusively (analogously), inclusively (with corresponding statistical weights) or in a mixed mode. There are several options for the geometry, with "extended" or ROOT-based [18] modes as the most commonly used ones. Photoproduced muons are included in the MARS code in two ways:

- An exclusive muon generator based on the Weizsäcker-Williams approximation using algorithms based on the work of [19]. Only coherent photomuon production is simulated. This generator is used as the default generator for muon production.

- An inclusive muon generator based on the calculation of the lowest-order Born approximation in [16,17] for targets of arbitrary mass, spin and form factor as well as arbitrary final states.

Both models give practically identical results for photon energies larger than 10 GeV. At lower energies, a precise description of the nuclear form factors becomes important. MARS supports two options for the description of the nuclear density for the inclusive muon generator: the original Tsai power-law mode and a symmetrized Fermi function. Angular and momentum distributions of muons produced by bremsstrahlung photons of 18 GeV electrons in copper simulated with the inclusive and the exclusive generator are in close agreement. The Weizsäcker-Williams approximation is therefore adequate for the benchmark in question.

*GEANT4.10*
The GEANT4 toolkit is the successor of the series of GEANT programs for geometry and tracking developed at CERN. It is based on object-oriented software technology. GEANT4 represents a set of software tools from which the user needs to program his own application. The implementation of muon pair production is described in [19] and is based on the work in [20].

The geometry of the experiment has been modeled with the three codes using information from [1,21]. A consistent geometry was defined at the SATIF-12 workshop for the three codes. Vertical views of the geometry are shown in Figure 1 (FLUKA), Figure 2 (MARS15) and Figure 3 (GEANT4). Figure 1 also indicates the location of the beamdump and the gaps in which the detectors were placed in the experiment. The electron beam is coming from the left and hits the beamdump.

**Figure 1: Geometrical model of the experimental setup using FLUKA2011**

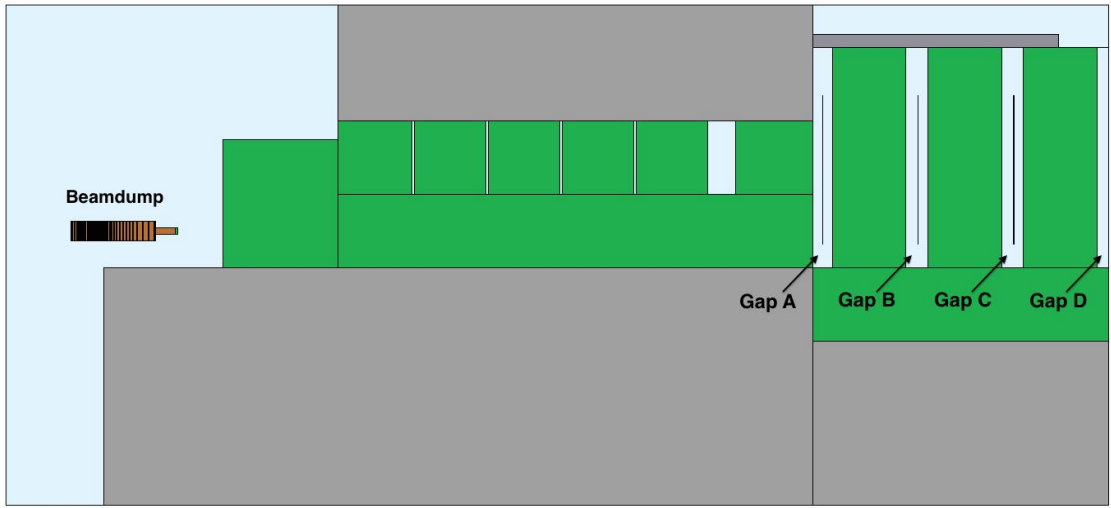

**Figure 2: Geometrical model of the experimental setup using MARS15**

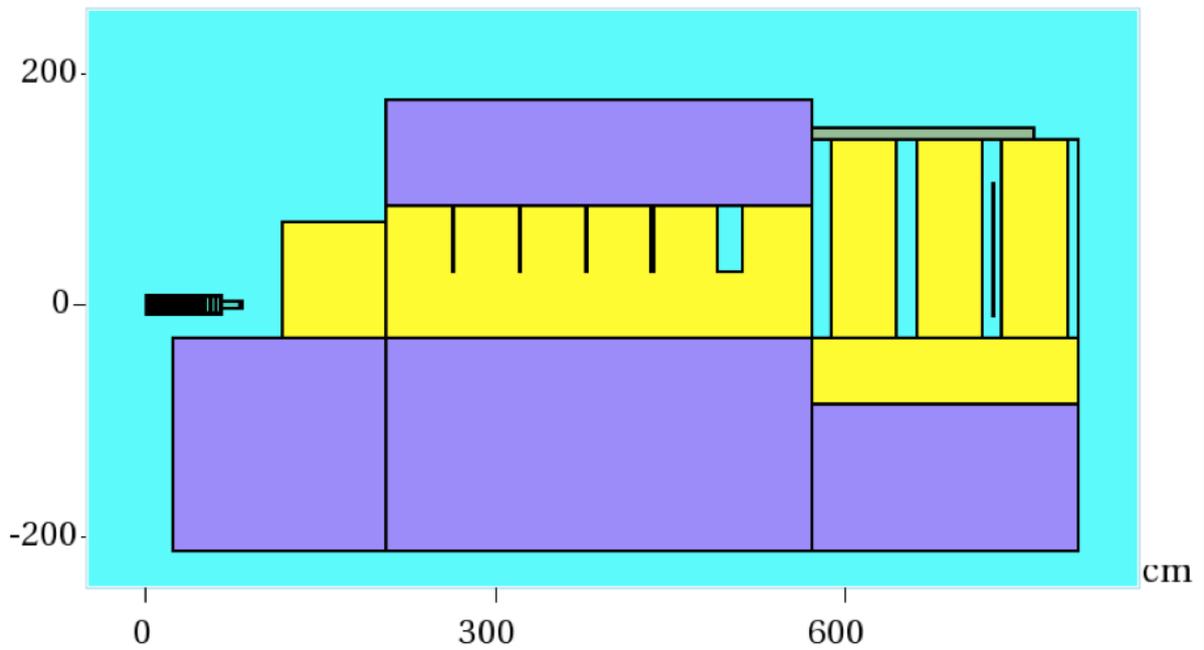

**Figure 3: Geometrical model of the experimental setup using GEANT4.10**

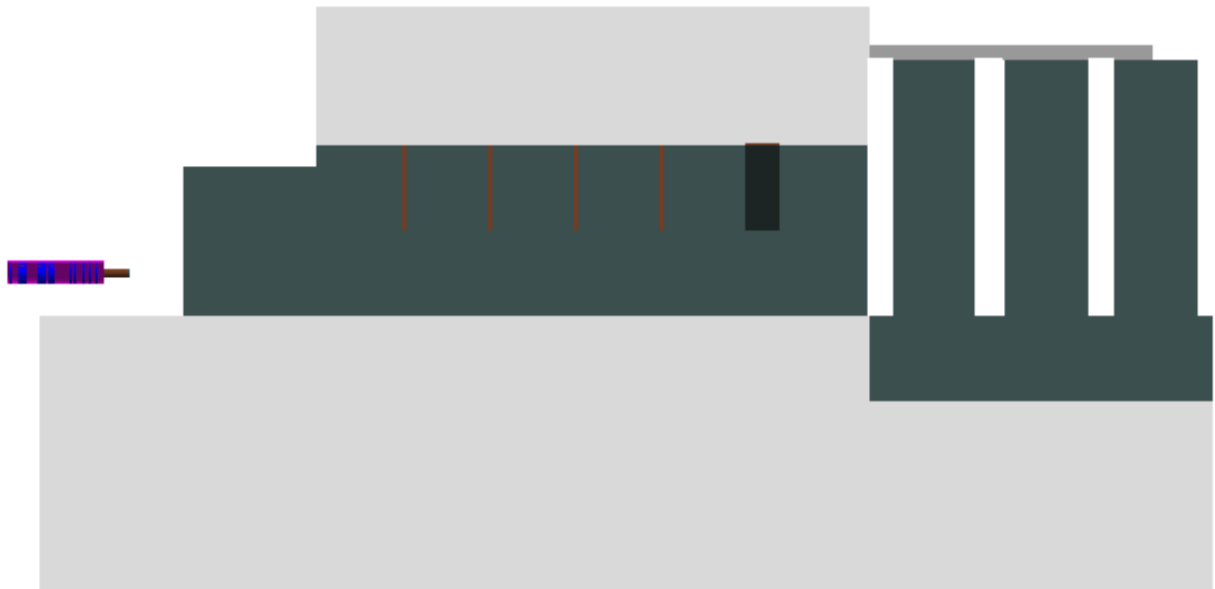

**Scoring and simulation parameters**

In order to score the results with the different transport codes and compare with the experimental results, the following scorers were defined:

- The **muon fluence** in the 4 gaps normalized to the integrated electron charge on the beam dump (in muons/cm$^2$/Coulomb)

- The **absorbed dose** in the 4 gaps normalized to the integrated electron charge on the beam dump (in rad/Coulomb). To simulate the dose deposition in the thermoluminescent dosimeters, thin layers of LiF (500µm thickness) are placed in each gap.

- Several double-differential scorers in **energy and angle** for muons crossing the copper-water intersections over approximately 6 radiation lengths in the beamdump allow to cross check the implementation of muon photoproduction in the different codes.

For MARS, scoring distributions are obtained via post-processing using PAW [22]. FLUKA scoring distributions are obtained from the built-in scorers and post-processing is done using FLAIR [15]. With GEANT4, a mixture of built-in scoring and ROOT histograms [18] is used.

The following simulation parameters and configuration options were used in the simulations:

1. MARS (used version: MARS15 (2016)):

- Generation and transport thresholds for electrons, positrons and gammas are set at 2 GeV in the beamdump, and at the following values elsewhere: 1E-9 MeV for neutrons, 1 MeV for muons and charged hadrons, 0.01 MeV for photons, and 0.1 MeV for electrons and positrons
- The default (exclusive) photo-muon generator is used

2. FLUKA (used version: FLUKA2011.2c.3):

- Defaults for precision simulations are used
- Production and transport thresholds for electrons and photons are set to 100 keV and 10 keV, respectively
- Full simulation of muon nuclear interactions and production of secondary hadrons switched on
- Production of secondaries for muons and charged hadrons switched on (100 keV threshold)

3. GEANT4 (used version: GEANT4.10.2p02):

- Basic physics list with quark gluon string and Bertini models is used, with parameters for electromagnetic physics tuned for high precision
- Additional process for gamma conversion to muons switched on for photons
- Additional process for muon-nucleus interactions switched on for muons
- Range threshold for gamma, electron, positron, proton: 700μm

**Status at SATIF-12 and progress since then**

At the last SATIF workshop, first results were presented from FLUKA simulations [4]. At that time, the composition of the shielding material was still unclear, and therefore two simulations were performed for different types of steel with a density of $\rho=7.0$ g/cm$^3$ and $\rho=7.6$ g/cm$^3$. In the meantime, some of the shielding blocks used in the experiment were identified at SLAC, and measurements were performed to obtain the density and elemental composition [23]. A density of $\rho=7.6$ g/cm$^3$ was found. The elemental composition of the steel is reported in Table 1 and Figure 4. In addition to the material composition and density for the shielding steel, small updates in the geometry were applied to the simulation:

- A void for a second beamline was introduced into the shielding structures
- The shielding for gaps A, B and C with lead blocks was added

While these geometrical updates were partly already included in the simulations, they were not yet used for the FLUKA results on fluence and dose presented in [4].

**Table 1: Composition of the steel used for the simulation of the shielding structures (density ρ=7.6g/cm³)**

| Elements | C | N | Si | S | O |
|---|---|---|---|---|---|
| Mass fraction [%] | 0.647 | 2.5E-3 | 0.1625 | 0.023 | 1.3 |
| Elements | Pb | Cd | Cr | Ti | Cu |
| Mass fraction [%] | 0.02859 | 1.86E-3 | 2.769E-2 | 0.3022 | 0.0708 |
| Elements | Zn | Zr | Sn | Mn | Co |
| Mass fraction [%] | 2.98E-2 | 7.87E-3 | 9.976E-3 | 0.675 | 0.2037 |
| Elements | Ni | Mo | Nb | Fe | |
| Mass fraction [%] | 0.02 | 0.0659 | 0.1238 | 96.297317 | |

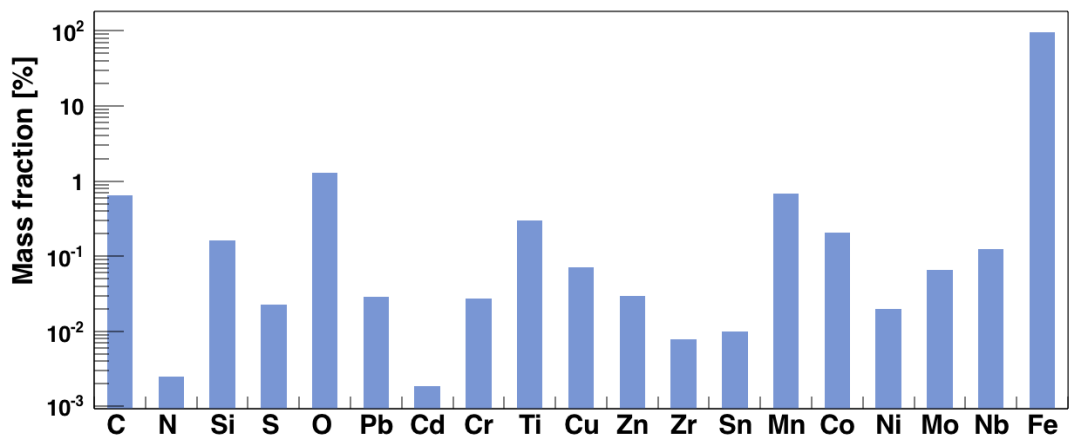

Figure 4: Composition of the steel used for the simulation of the shielding structures (density ρ=7.6g/cm³)

**Results**

Figures 5-8 show the muon fluence registered in the four gaps for data (black triangles) compared to the FLUKA simulation (red dots), the MARS15 simulation (blue squares) and the GEANT4 simulation (green triangles). Only simulated data points with statistical uncertainty < 30% were kept. All codes represent the experimental data reasonably well, however there are small differences:

- The MARS15 points reproduce the shape of the data points very well, but fall systematically slightly lower
- FLUKA points have a less steep slope than the data, and are therefore lower at smaller polar angles and higher at angles above 40 – 80 mrad
- GEANT4 points tend to fall between the MARS15 and the FLUKA results, with a tendency to be closer to the FLUKA points at gaps C and D above 40 mrad

It is interesting to note that the trends for MARS15 and FLUKA data are already hinted at in the plots in [12]. While the systematic shift of the MARS15 data could possibly be explained by air gaps in the shielding which are unaccounted for in the simulation geometry and reduce the effective amount of shielding, the different slope of the FLUKA data could be caused by missing nuclear form factors in the treatment of multiple scattering by muons.

**Figure 5: Results for muon fluence in Gap A**

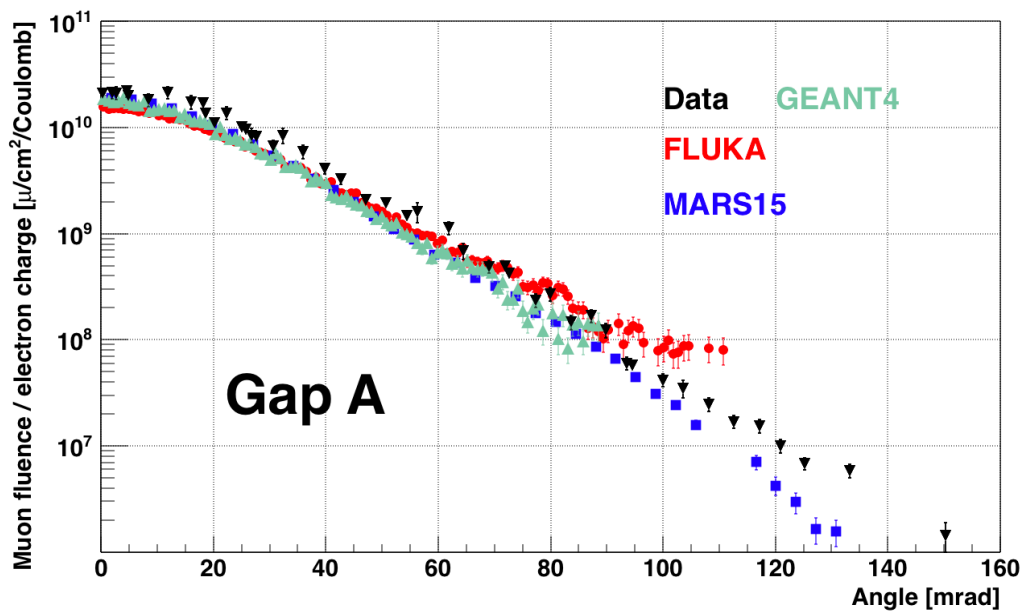

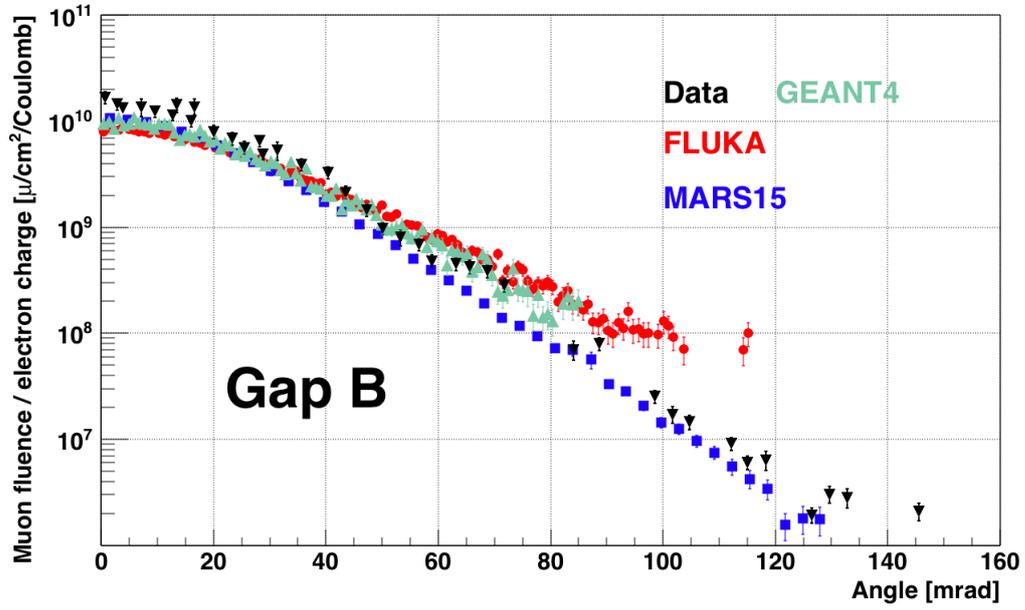

Figure 6: Results for muon fluence in Gap B

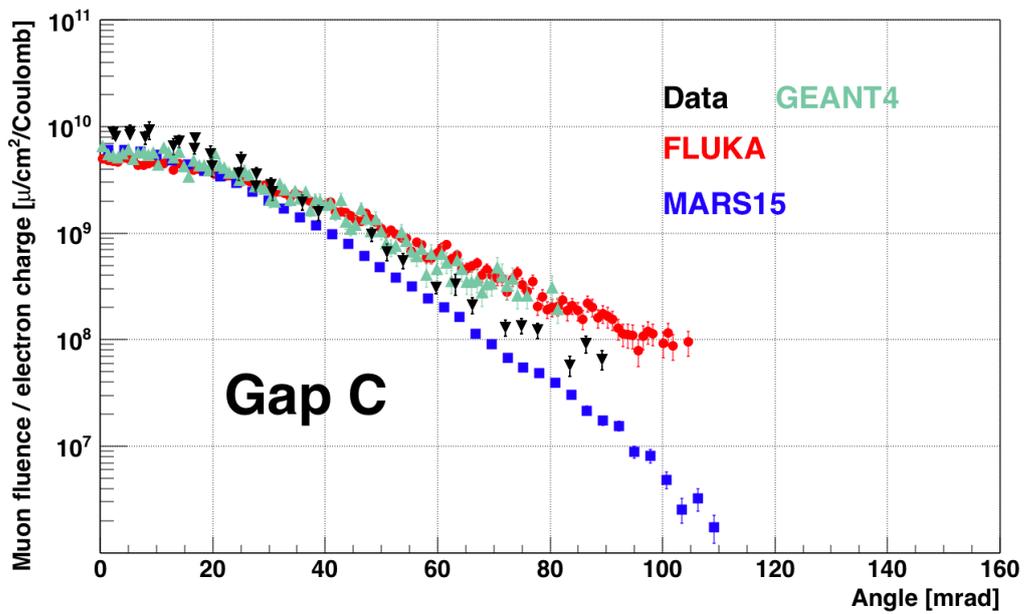

Figure 7: Results for muon fluence in Gap C

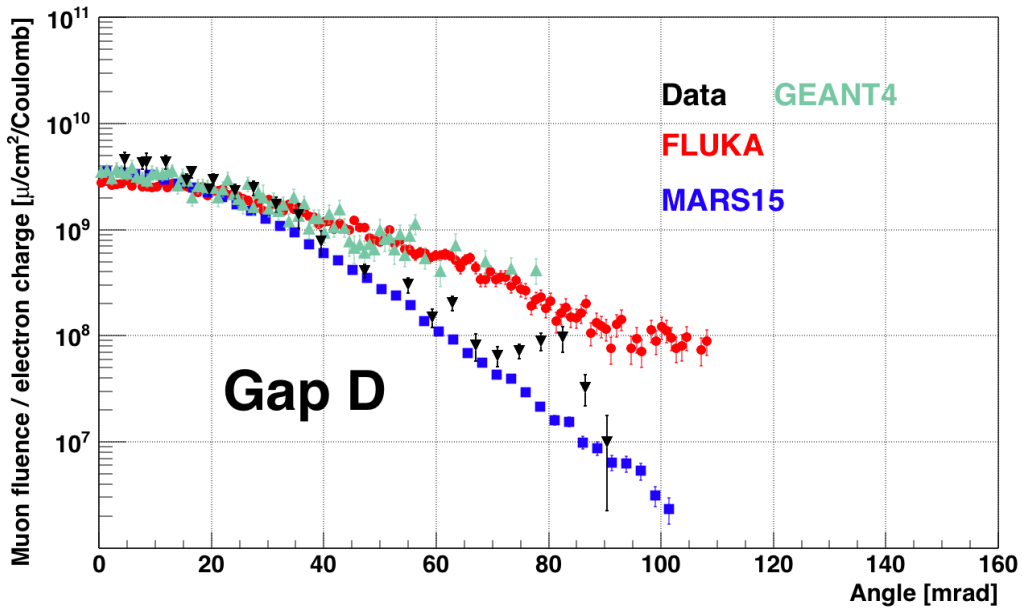

Figure 8: Results for muon fluence in Gap D

Figures 9-12 show the absorbed dose as function of the polar angle registered in the four gaps for data (black triangles) compared to the FLUKA simulation (red dots), the MARS15 simulation (blue squares) and the GEANT4 simulation (green triangles). Again, only simulated data points with statistical uncertainty < 30% were kept. The codes are reasonably close to the experimental data points, except for gap D, where especially at angles larger than 40 mrad the data points are significantly larger than the simulations. This is probably due to the missing lead shielding for this gap in the experiment, which allows backscattered neutrons from walls, floor and ceiling (and even air) to contribute to the dose. Since at the moment, the simulation geometry ends right after gap D, these backscattering effects are not taken into account in the simulations. For gaps A-C, visible agreement below 40 mrad is very good between simulations and experimental data, at larger angles, the different slopes of MARS15 and FLUKA points lead to a situation in which the FLUKA points are above and the MARS15 points are below the experimental results.

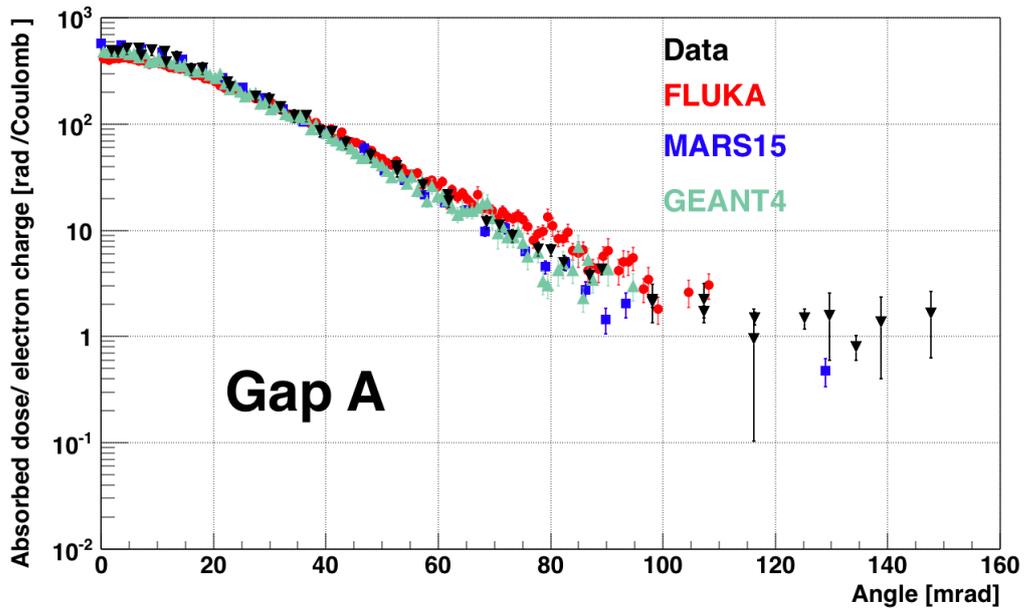

Figure 9: Results for absorbed dose in Gap A

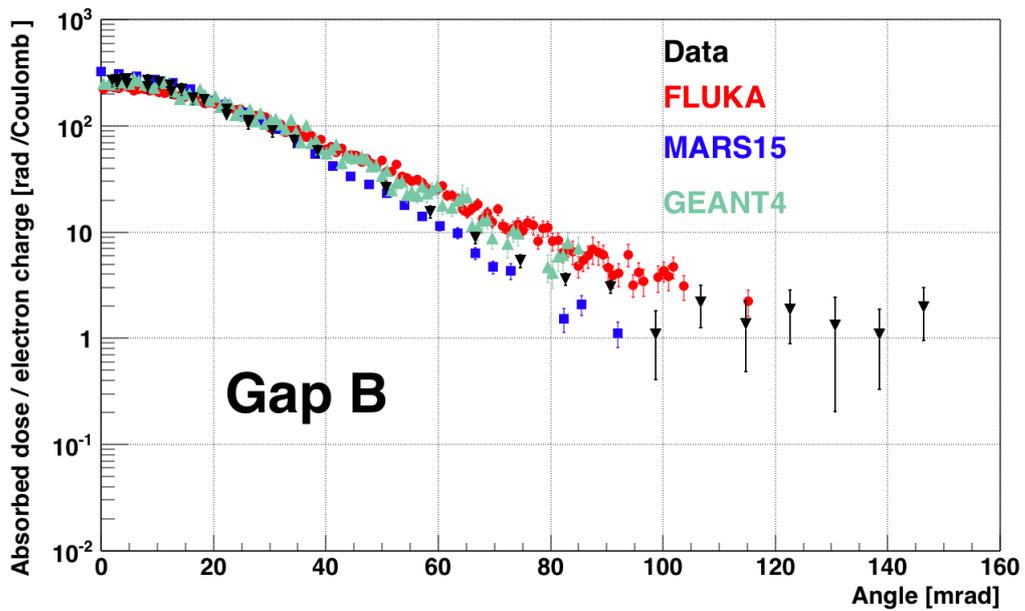

Figure 10: Results for absorbed dose in Gap B

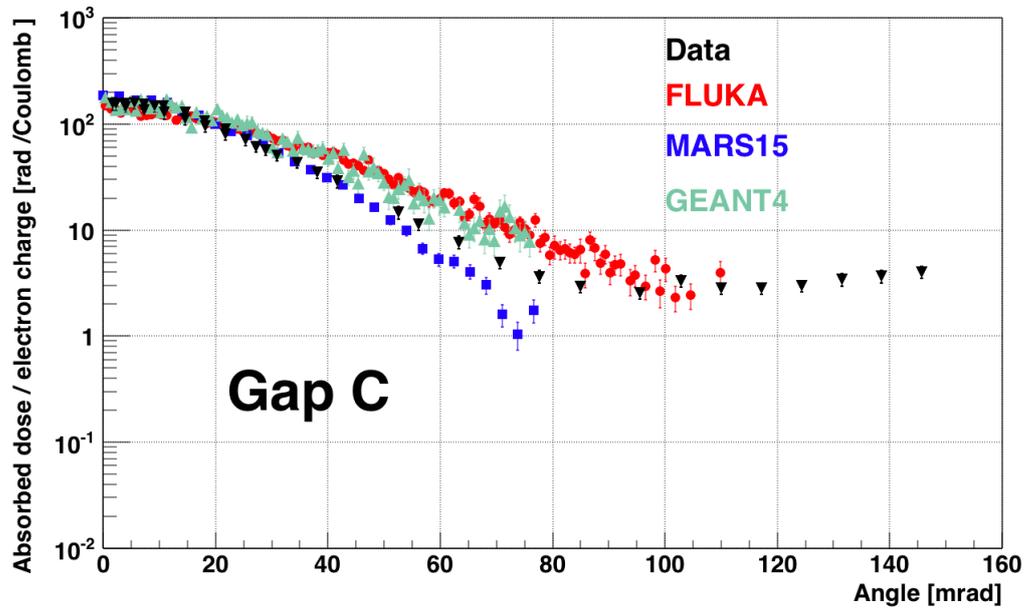

Figure 11: Results for absorbed dose in Gap C

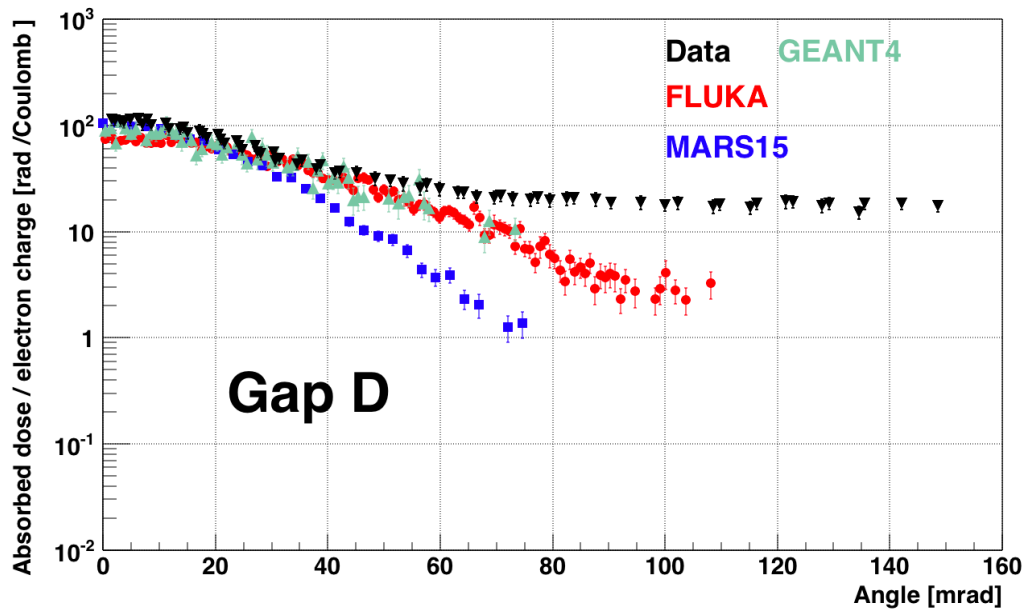

Figure 12: Results for absorbed dose in Gap D

**Summary and conclusions**

Based on the experimental results on muon production by an 18 GeV e⁻ beam hitting a copper-water target reported by Nelson, Kase and Svensson, the Monte Carlo transport codes MARS15, FLUKA and GEANT4 have been used to model the experimental conditions. First preliminary results on muon fluence and absorbed dose have been produced with the three codes using updated material and geometry definitions in the simulations. The agreement between the simulated results and the experimental values is quite promising. Dedicated consistency checks using double-differential distributions of muon fluence at several positions in the beamdump-target will allow to investigate more carefully the production and transport of photo-produced muons in the different simulation programs.

**Acknowledgements**

The work of Fermilab authors was supported by Fermi Research Alliance, LLC under contract DE-AC02-07CH11359 with the U.S. Department of Energy.